\documentstyle[prd,aps,preprint,epsfig]{revtex}
\begin{document}
\tighten
\draft
\title{Quark-antiquark potential in the analytic approach to QCD}
\author{A.\ V.\ Nesterenko\thanks{Electronic address:
nesterav@thsun1.jinr.ru}}
\address{Physical Department, Moscow State University, Moscow,
119899, Russia}
\date{March 1, 2000}
\maketitle
\begin{abstract}
The quark-antiquark potential is constructed by making use of a new
analytic running coupling in QCD. This running coupling arises under
``analytization'' of the renormalization group equation. The rising
behavior of the quark-antiquark potential at large distances, which
provides the quark confinement, is shown explicitly. At small
distances, the standard behavior of this potential originating in the
QCD asymptotic freedom is revealed. The higher loop corrections and
the scheme dependence of the approach are briefly discussed.
\end{abstract}

\section{Introduction}
     The description of quark dynamics inside hadrons remains an
actual problem of elementary particle theory. The asymptotic freedom
in quantum chromodynamics (QCD) enables one to investigate the quark
interactions at small distances by making use of standard
perturbation theory. The quark dynamics at large distances (the
confinement region) lies beyond such calculations. For this purpose
other approaches are used: phenomenological potential models
\cite{Phenom}, string models \cite{Strings}, bags models \cite{Bags},
lattice calculations \cite{Lattice}, the explicit account of
nontrivial QCD vacuum structure~\cite{Brambilla}, and variational
perturbation theory~\cite{ILS1}.

     Recently Shirkov and Solovtsov proposed a new analytic approach
to QCD \cite{ShSol1}. Its basic idea is the explicit imposition of
the causality condition, which implies the requirement of the
analyticity in the $Q^2$ variable for the relevant physical
quantities. The essential merits of this approach are the following:
absence of unphysical singularities at any loop level, stability in
the infrared (IR) region, stability with respect to loop corrections,
and extremely weak scheme dependence. The analytic approach has been
applied successfully to such problems as the $\tau$ lepton decays,
$e^+e^-$-annihilation into hadrons, sum rules (see \cite{ShSol2} and
references therein).

     In Refs.\ \cite{MyDipl,L99} the analytic approach has been
employed to the solution of the renormalization group (RG) equation.
The analyticity requirement was imposed on the RG equation itself,
before deriving its solution. Solving the RG equation, ``analytized''
(i.e., requiring analyticity) in the above-mentioned way, one gets,
at one-loop level, a new analytic running coupling, which possesses
practically the same appealing features as the Shirkov-Solovtsov
running coupling \cite{ShSol1} does. An essential distinction, that
will play a crucial role in the present paper, is the IR singularity
of the new analytic running coupling at the point $q^2=0$. It should
be stressed here that such a behavior of the invariant charge is in a
complete agreement with the Schwinger-Dyson equations, and, as it
will be demonstrated in Sec.\ III, provides the quark confinement
(see Sec.\ II for the details).

     In this paper we shall adhere to the model
\cite{Brambilla,Fourier} of obtaining the quark-antiquark ($q\bar q$)
potential by the Fourier transformation of the running coupling.
However, the perturbative running coupling ${\alpha}_{\text{s}}(q^2)$
does not enable one to obtain the rising $q\bar q$ potential without
invoking additional assumptions \cite{Fourier}.

     The objective of this paper is to construct the quark-antiquark
potential by making use of the new analytic running coupling. This
potential proves to be rising at large distances (i.e., providing the
quark confinement) and, at the same time, it incorporates the
asymptotic freedom at small distances. It is essential that for
obtaining this potential {\it no additional assumptions}, lying
beyond the standard RG method in the quantum field theory and the
analyticity requirement, will be used.

     The layout of the paper is as follows. In Sec.\ II the
derivation of the new analytic running coupling is presented and its
properties are briefly discussed. In Sec.\ III the quark-antiquark
potential, generated by the new analytic running coupling, is derived
by making use of the Fourier transformation. Further, the asymptotic
behavior of the potential at large and small distances is
investigated. In Sec.\ IV the higher loop corrections and the scheme
dependence of the potential are discussed briefly. For practical
purposes, a simple approximate formula for the potential is proposed
which interpolates its infrared and ultraviolet asymptotics. This
formula is compared with the phenomenological Cornell potential.
Proceeding from this, an estimation of the QCD parameter $\Lambda $
is obtained. In the Conclusion (Sec.\ V) the obtained results are
formulated in a compact way, and the further studies in this approach
are outlined.

\section{A new model for the QCD analytic running coupling}
     In the analytic approach to QCD, proposed by Shirkov and
Solovtsov \cite{ShSol1}, the basic idea is the explicit imposing of
the causality condition, which implies the requirement of the
analyticity in the $Q^2$ variable for the relevant physical
quantities. Later this idea was applied to the ``analytization''
(i.e., the procedure of analyticity requirement) of the perturbative
series when calculating the QCD observables \cite{ShSol2}. The
results turned out to be quite encouraging. As was mentioned in the
Introduction, the analytization of the perturbative series leads to
the elimination of the unphysical singularities, to the higher loop
correction stability and to a weak scheme dependence. However, the
$Q^2$-evolution of some QCD observables (for instance, the structure
function moments) is intimately tied with the solution of the RG
equation. Our task here is to involve the analytization procedure
into the RG formalism more profoundly.

     Let us consider the RG equation of a quite general form for a
quantity ${\sf A}(Q^2)$ (it may be, for example, the gluon
propagator, or the structure function moment). At the one-loop level
this equation reads
\begin{equation}
\label{StdRGEqn}
\frac{d\ln{\sf A}(Q^2)}{d\ln Q^2} = \gamma\,
\widetilde{\alpha}^{(1)}_{\text{s}}(Q^2),
\end{equation}
where $\gamma$ is the corresponding anomalous dimension (the negative
noninteger number in the general case),
$\widetilde{\alpha}^{(1)}_{\text{s}}(Q^2) = 1/\ln(Q^2/\Lambda^2)$ is
the one-loop perturbative running coupling. The solution of Eq.
(\ref{StdRGEqn}) can be written in the form
\begin{equation}
\label{StdRGEqnSol}
{\sf A}(Q^2) = {\sf A}(Q^2_0)
\left[\frac{\widetilde{\alpha}^{(1)}_{\text{s}}
(Q^2_0)}{\widetilde{\alpha}^{(1)}_{\text{s}}(Q^2)}\right]^{\gamma}.
\end{equation}
From here it follows immediately, that this solution has unphysical
singularities in the physical region $Q^2>0$. However, in many
interesting cases mentioned above, the quantity ${\sf A}(Q^2)$ must
have correct analytic properties in the $Q^2$ variable (namely,
there is the only cutoff $Q^2\le 0$). One can demonstrate this
proceeding from the first principles. So, for the gluon propagator
this assertion follows from the causal K\"all\'en-Lehmann
representation (see, e.g., \cite{BgSh}), and for the structure
function moments this is a consequence of the Deser-Gilbert-Sudarshan
integral representation\footnote{In the most general case this
follows from the Jost-Lehmann-Dyson integral representation for the
structure function, but the detailed discussion of this point is
beyond the scope of this paper.} (see, e.g., \cite{Wetz}). Thus, we
come to a contradiction.

     The point, which is crucial to our consideration, is the
following. The RG equation in the form (\ref{StdRGEqn}) involves, in
fact, a contradiction. The left-hand side of this equation has no
unphysical singularities in the $Q^2>0$ region, while its right-hand
side has pole-type singularity at the point $Q^2=\Lambda^2$. The
account of the higher loop contributions just introduces the
additional unphysical singularities of the cut type in the physical
region $Q^2>0$ and hence does not solve the problem.

     In order to avoid this contradiction, we propose to use the
following method \cite{MyDipl,L99}. Before solving the RG equation
(\ref{StdRGEqn}) one should analytize its right-hand side as a whole.
This prescription leads to the analytized RG equation, which, at the
one-loop level, takes the form
\begin{equation}
\label{ARGEqn}
\frac{d\ln{\sf A}(Q^2)}{d\ln Q^2} = \gamma\,
\widetilde{\alpha}^{(1)}_{\text{an}}(Q^2),
\end{equation}
where $\widetilde{\alpha}^{(1)}_{\text{an}}(Q^2) =
1/\ln(Q^2/\Lambda^2) + 1/(1-Q^2/\Lambda^2)$ is the perturbative
running coupling analytized by making use of the Shirkov-Solovtsov
prescription \cite{ShSol1}.\footnote{It is worth noting here that
there is no consistent way for analytizing the RG equation with the
$\beta$-function for the invariant charge.} The solution of Eq.\
(\ref{ARGEqn}) can be presented in the form
\begin{equation}
\label{ARGEqnSol}
{\sf A}(Q^2)={\sf A}(Q^2_0)
\left[
\frac{^{\text{N}}\widetilde{\alpha}^{(1)}_{\text{an}}(Q^2_0)}
{^{\text{N}}\widetilde{\alpha}^{(1)}_{\text{an}}(Q^2)}
\right]^{\gamma},
\end{equation}
where
\begin{equation}
\label{NARC}
^{\text{N}}\widetilde{\alpha}^{(1)}_{\text{an}}(Q^2)
= \frac{z-1}{z\ln z},\quad z=\frac{Q^2}{\Lambda^2}.
\end{equation}
Comparing the solution (\ref{ARGEqnSol}) with Eq.\
(\ref{StdRGEqnSol}) one infers that
$^{\text{N}}\alpha^{(1)}_{\text{an}}(Q^2)$ should be treated as a new
one-loop analytic running coupling. Really, it possesses the same
properties, as the one-loop running coupling analytized through the
Shirkov-Solovtsov procedure. Namely, the new running coupling has the
standard asymptotic behavior at $z\to\infty$ and it has no unphysical
singularities in the $Q^2>0$ region. The latter follows directly from
the causal representation of the K\"all\'en-Lehmann type, that holds
for $^{\text{N}}\widetilde{\alpha}^{(1)}_{\text{an}}(Q^2)$:
\begin{equation}
\label{NARCIntRep}
^{\text{N}}\widetilde{\alpha}^{(1)}_{\text{an}}(Q^2) =
\int_{0}^{\infty}
\frac{^{\text{N}}\rho(\sigma)}{\sigma+z}d\sigma,\quad
^{\text{N}}\rho(\sigma)=
\left(1+\frac{1}{\sigma}\right)\frac{1}{\ln^2\sigma+\pi^2}.
\end{equation}

     The distinctive feature of the new running coupling, which will
play the crucial role in the framework of our consideration, is its
singularity at the point $Q^2=0$. It is worth noting that such a
behavior of the invariant charge is in a complete agreement with the
Schwinger-Dyson equations (see discussion in Ref.\ \cite{Alek}), and,
as it will be demonstrated in the next section, provides the quark
confinement.

     Summarizing all stated above, we propose the following model
for the analytic running coupling. We define the new analytic running
coupling $^{\text{N}}\alpha_{\text{an}}(Q^2)$ as the solution of the
analytized RG equation at the respective loop level. Here one has to
choose the anomalous dimensions in such a way that the solution of
the standard RG equation is the perturbative running coupling at the
loop level considered.\footnote{This choice is the following:
$\gamma = \gamma_0 = -1$; $\gamma_i=0,\;i\ge 1$, where $\gamma_i$ is
the coefficient by the $(i+1)$th power of the perturbative running
coupling on the right-hand side of Eq.\ (\ref{StdRGEqn}).} Thus, at
the one-loop level, the new analytic running coupling has the form
\cite{MyDipl,L99}
\begin{equation}
\label{NARCDef}
^{\text{N}}\alpha^{(1)}_{\text{an}}(Q^2) =
\frac{4\pi}{\beta_0}\frac{z-1}{z\ln z},\quad
z = \frac{Q^2}{\Lambda^2},
\end{equation}
where $\beta_0=11-2\,n_f/3$ is the first coefficient of the
$\beta$-function. At the higher loop levels there is only the
integral representation for $^{\text{N}}\alpha_{\text{an}}(Q^2)$. So,
at the $i$-loop level we have
\begin{equation}
\label{NARCDefHL}
^{\text{N}}\alpha^{(i)}_{\text{an}}(Q^2) =
~^{\text{N}}\alpha^{(i)}_{\text{an}}(Q^2_0)\frac{z_0}{z}
\exp\left[\int_{0}^{\infty}\frac{\rho^{(i)}(\sigma)}{\sigma}
\ln\left(\frac{\sigma+z}{\sigma+z_0}\right)d\sigma\right],
\end{equation}
where $\rho^{(i)}(\sigma) = \left[
\widetilde{\alpha}^{(i)}_{\text{s}}(-\sigma-i\varepsilon) -
\widetilde{\alpha}^{(i)}_{\text{s}}(-\sigma+i\varepsilon)
\right]/(2\pi i)$ is the spectral density, and $z_0=Q^2_0/\Lambda^2$
is the normalization point.

\begin{figure}
\noindent
\centerline{\epsfig{file=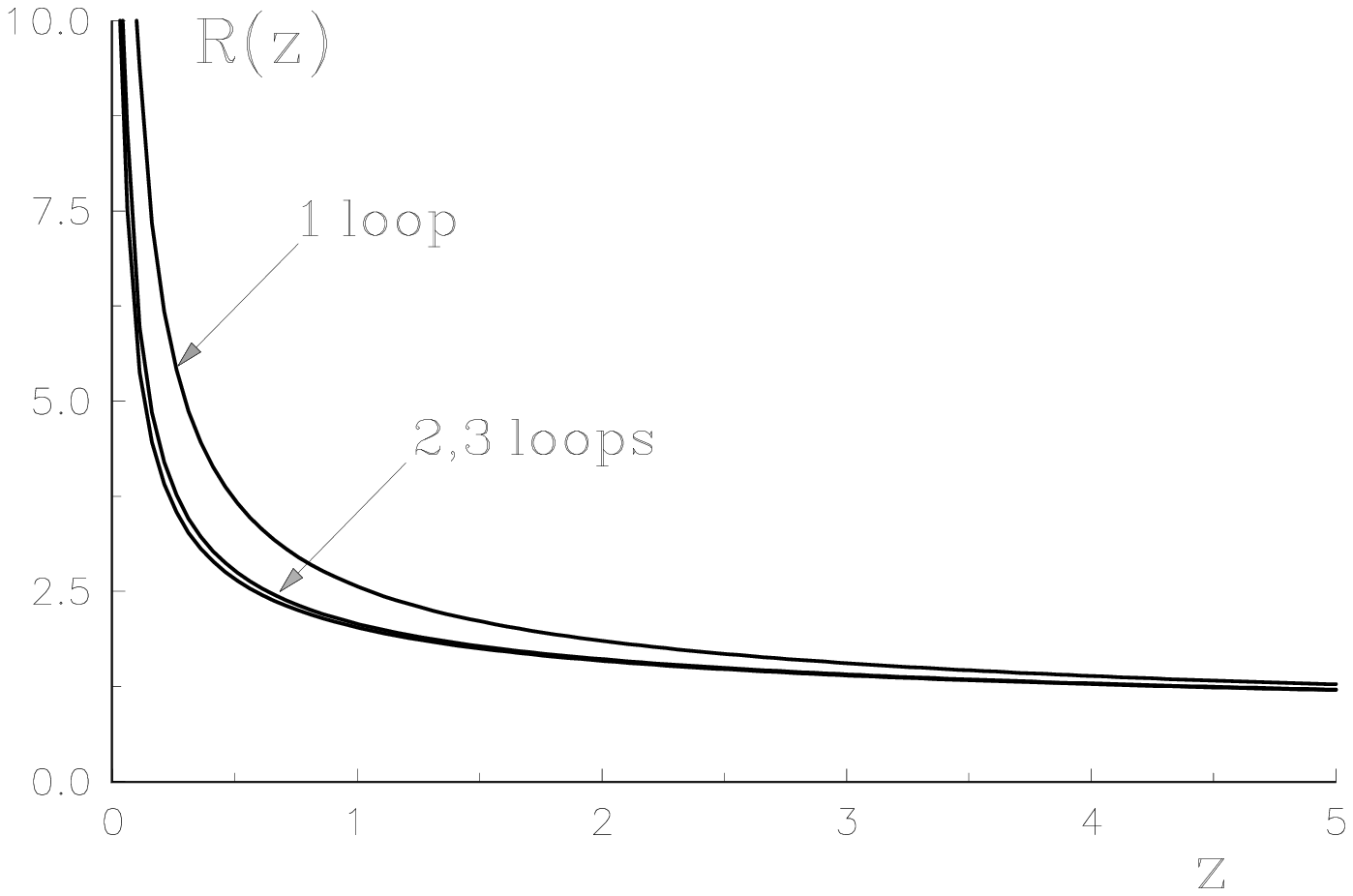, width=125mm}}
\caption{The normalized new analytic running coupling $R(z) =
~^{\text{N}}\widetilde{\alpha}_{\text{an}}(Q^2)
/~\!^{\text{N}}\widetilde{\alpha}_{\text{an}}(Q^2_0)$ at the
one-, two-, and three-loop levels. The normalization point is
$Q^2_0=10\Lambda^2$, $\,z=Q^2/\Lambda^2$.}
\label{NARCGraph}
\end{figure}

     Figure \ref{NARCGraph} shows the new analytic running coupling
computed at the one-, two-, and three-loop levels. It is clear from
this figure that our analytic running coupling possesses the higher
loop stability. Moreover, it can be shown that the singularity of the
new analytic running coupling at the point $Q^2=0$ is of the
universal type at any loop level. This is clear from the following
simple consideration. When $z\to 0$ the basic contribution into Eq.\
(\ref{NARCDefHL}) affords the integration over the small $\sigma$
region. The spectral density $\rho^{(i)}(\sigma)$ at any loop level
has the same limit when $\sigma\to 0$: $\rho^{(i)}(\sigma) \to
\rho^{(1)}(\sigma) = (\ln^2 \sigma + \pi^2)^{-1}$ \cite{ShSol2}.
Hence the new analytic running coupling (\ref{NARCDefHL}) has the
unique behavior when $z\to 0$.

\section{Quark-antiquark potential generated by the new analytic
         running coupling}
     Here we are going to use the new analytic running coupling for
obtaining the interquark potential. We proceed from the standard
expression \cite{Brambilla,Fourier} for the $q\bar q$ potential in
terms of the running coupling $\alpha(q^2)$,
\begin{equation}
\label{VrGen}
V(r) = -\frac{16\pi}{3} \int_{0}^{\infty}\frac{\alpha(q^2)}{q^2}
\frac{e^{i{\bf qr}}}{(2\pi)^3}\,d{\bf q}.
\end{equation}

     For the construction of the new interquark potential $^NV(r)$ we
shall use the new analytic running coupling~(\ref{NARCDef})
\begin{equation}
\label{ARCDef}
^{\text{N}}\alpha_{\text{an}}(Q^2) =
\frac{4\pi}{\beta_0}
\frac{z-1}{z\ln z}, \quad z=\frac{Q^2}{\Lambda ^2}.
\end{equation}
Upon the integration over the angular variables
and the substitution $q/\Lambda\to q, \,r\Lambda\to R$ in Eq.\
(\ref{VrGen}), one gets
\begin{equation}
\label{NVrDef}
^NV(r)=-\frac{32}{3\beta_0}\Lambda\cdot\widetilde{V}(R),
\quad R=\Lambda r,
\end{equation}
where
\begin{equation}
\label{NVtrDef}
\widetilde{V}(R) = \int_{0}^{\infty}\frac{q^2-1}{q^2\ln q^2}
\frac{\sin(qR)}{qR}\,dq
\end{equation}
is the dimensionless potential.

     In order to perform the integration in Eq.\ (\ref{NVtrDef}) we
consider the auxiliary function
\begin{equation}
\label{IDef}
I(n, R) = \lim_{a\to 0+}\int_{0}^{\infty} \frac{q^{n-1}}{\ln(a+q^2)}
\sin(qR) \,dq.
\end{equation}
Here the parameter $a$ is introduced for shifting the origin of the
cut along the imaginary axis {\rm Im}~$q$. It is obvious that
\begin{equation}
\label{VtRDef}
\widetilde{V}(R) = \frac{1}{R}\left[I(0, R) - I(-2, R)\right].
\end{equation}
For even $n$ the integrand in Eq.\ (\ref{IDef}) is an even function
of~$q$. Therefore
\begin{equation}
\label{JDef}
I(n, R)=\frac{1}{2}\: \mbox{\rm Im}\, \lim_{a\to 0+}J(n, a, R),
\end{equation}
where
\begin{equation}
\label{FDef}
J(n,a,R)={\cal P}\!\!\int_{-\infty}^{\infty}\!\!F(q)\,dq,\quad
F(q)=\frac{q^{n-1} e^{iqR}}{\ln(a+q^2)}.
\end{equation}
The sign ${\cal P}$ means the principal value of the integral.

\begin{figure}
\noindent
\centerline{\epsfig{file=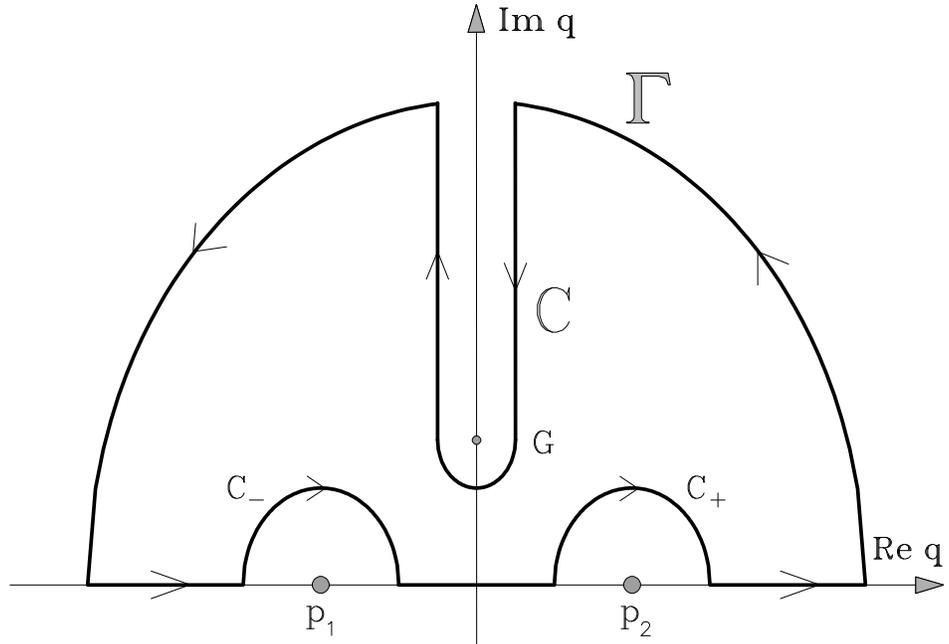, width=125mm}}
\caption{The integration contour in the complex $q$-plane.
The notations are: $p_1=-\protect\sqrt{1-a},\;$
$p_2=\protect\sqrt{1-a},\;$ G= $i \protect\sqrt{a}$.}
\label{Contour}
\end{figure}

     The function $F(q)$ in Eq.\ (\ref{FDef}) has the cuts
$(-i\infty, -i\sqrt{a}]$, $\,[i\sqrt{a}, i\infty)$ and simple poles
at the points $q=\mp\sqrt{1-a}$. Let us consider the integral of the
function $F$ along the contour $\Gamma$ shown in Fig.\ \ref{Contour}.
The function $F(q)$ has no singularities inside the contour $\Gamma$,
therefore $\oint_{\Gamma}\!F\,dq=0$. Contribution to this integral of
the semicircle of infinitely large radius in upper half-plane (see
Fig.\ \ref{Contour}) vanishes. Performing the integration along the
two semicircles $c_-$ and $c_+$ of the vanishing radius and along the
cut $C$ on the imaginary axis, we obtain
\begin{eqnarray}
J(n,a,R) &=& i\pi \Biggl\{\frac{1}{2}\left[
(\sqrt{1-a})^{n-2}e^{iR\sqrt{1-a}}\right.\nonumber\\
&&+\left.(-\sqrt{1-a})^{n-2}e^{-iR\sqrt{1-a}}\right]
\nonumber\\
&&+ 2i^{n-2}
\int_{\sqrt{a}}^{\infty}\frac{x^{n-1}e^{-Rx}}{\ln^2(x^2-a)+\pi^2}\,dx
\Biggr\}.
\end{eqnarray}
Hence, for even $n$ the function $I(n,R)$ in Eq.\ (\ref{IDef}) takes
the form
\begin{equation}
\label{InDef}
I(n,R) = \pi \left[\frac{1}{2}\cos R -
(-1)^{n/2}{\cal N}(R,n)\right],
\end{equation}
where
\begin{equation}
\label{NDef}
{\cal N}(R,n) = \int_{0}^{\infty}
\frac{x^{n-1}e^{-Rx}}{\ln^2(x^2) + \pi^2}\, dx.
\end{equation}

     It is rather complicated to perform the integration in Eq.\
(\ref{NDef}) explicitly. Therefore we address the study of the
asymptotics. First of all, we would like to know whether the $q\bar
q$ potential $^NV(r)$ in Eq.\ (\ref{NVrDef}) provides the quark
confinement. For the investigation of the potential behavior at
large distances it is enough to consider the asymptotic of the
function ${\cal N}(R,n)$ in Eq.\ (\ref{NDef}) when $R\to\infty$.
This function can be represented in the following way
\begin{equation}
\label{NReDef}
{\cal N}(R,n) =
(-1)^n \frac{\partial^n}{\partial R^n} \int_{0}^{\infty}
\frac{e^{-Rx}}{x\left[\ln^2(x^2) + \pi^2\right]}\, dx.
\end{equation}
At large $R$ the basic contribution into Eq.\ (\ref{NReDef}) gives
the integration over the small $x$ region. Let us transform ${\cal
N}(R,n)$ identically:
\begin{equation}
\label{NDiff}
{\cal N}(R,n)=\frac{\partial^n}{\partial R^n}
\int_{0}^{\infty}\frac{(-1)^ne^{-Rx}}{4x(\ln^2 x+\pi^2)}
\left[1 + \frac{3L}{1+L}\right] dx,
\end{equation}
where $L=\pi^2/(4\ln^2x)$. Neglecting the second term in the square
brackets in Eq.\ (\ref{NDiff}), we use the formula (4.361.2) from
Ref.\ \cite{RGrad}:
\begin{equation}
\int_{0}^{\infty} \frac{e^{-\mu x}}{x(\ln^2 x+\pi^2)} dx =
e^\mu-\nu(\mu), \quad \mbox{\rm Re} \; \mu>0,
\end{equation}
where $\nu(\mu)$ is the so-called transcendental $\nu$-function
\cite{BatErd}:
\begin{equation}
\nu(\mu)=\int_{0}^{\infty} \frac{\mu^t dt}{\Gamma(t+1)}.
\end{equation}
Eventually, we obtain for $R \to\infty$,
\begin{equation}
\label{NInt}
{\cal N}(R,n)\simeq\frac{(-1)^n}{4}\left[
e^R-\nu(R)+\int_{0}^{-n}\frac{R^t dt}{\Gamma(t+1)}
\right].
\end{equation}

     Taking into account Eqs.\ (\ref{VtRDef}), (\ref{InDef}), and
(\ref{NInt}) one can present the quark-antiquark potential
(\ref{NVrDef}) at large $R$ in the following way:
\begin{equation}
\label{NVrAs}
^NV(r) \simeq \frac{8\pi}{3 \beta_0}\frac{\Lambda}{R}\left[
2\left(e^R-\nu(R)\right)+\int_{0}^{2}\frac{R^t dt}
{\Gamma(t+1)}\right].
\end{equation}
The behavior of the potential $^NV(r)$ at $r\to\infty$ is determined
by the last term in Eq.\ (\ref{NVrAs}).\footnote{\label{Asymp} It
follows directly from the asymptotic of $\nu(R)$ (see Ref.\
\cite{BatErd}), and from a simple reasoning. Really, if $R>0$ the
term $f(R)=e^R-\nu(R)$ is non-negative and $ f'(R)\le 0$. Hence,
$f(R)\to $~const when $R\to\infty$, and its contribution to $^NV(r)$
at large $R$ is of $1/R$-order.} Integration of this term by parts
gives
\begin{eqnarray}
\label{PartInt}
\int_{0}^{2}\frac{R^t dt}{\Gamma(t+1)}&=&
\frac{R^2}{2\ln R}\left[1 + 2 \sum\limits_{k=1}^{\infty}
\frac{f_k(2)}{\ln^k R} \right]\nonumber\\
&&-\frac{1}{\ln R} \left[1 + \sum\limits_{j=1}^{\infty}
\frac{f_j(0)}{\ln^j R} \right],
\end{eqnarray}
where
\begin{equation}
f_n(t)=\left.\frac{d^n}{d s^n}\frac{(-1)^n}{\Gamma(s+1)}
\right|_{s=t}.
\end{equation}

     In the limit $R\to\infty$, Eq.\ (\ref{PartInt}) takes the form
\begin{equation}
\int_{0}^{2}\frac{R^t dt}{\Gamma(t+1)} = \frac{R^2}{2\ln R}.
\end{equation}
Therefore the quark-antiquark potential $^NV(r)$ proves to be rising
at large distances
\begin{equation}
\label{NVrAsInf}
^NV(r) \simeq \frac{8\pi}{3\beta_0}\Lambda\cdot\frac{1}{2}
\frac{\Lambda r}{\ln(\Lambda r)}, \quad r \to \infty.
\end{equation}

     Thus the new analytic running coupling
$^{\text{N}}{\alpha}_{\text{an}}(q^2)$ [see Eq.\ (\ref{NARCDef})]
leads to the rising quark-antiquark potential $^NV(r)$ which can, in
principle, describe the quark confinement.

     It is important to point out that the behavior of the potential
$^NV(r)$ when $r\to 0$ has the standard form determined by the
asymptotic freedom (see, e.g., Ref.\ \cite{Fourier}),
\begin{equation}
\label{NVrAsOrig}
^NV(r) \simeq \frac{8\pi}{3\beta_0}\Lambda\cdot\frac{1}
{\Lambda r \ln(\Lambda r)}, \quad r \to 0.
\end{equation}
Unfortunately, it is impossible to obtain the explicit dependence
$^NV(r)$ for the whole region $0<r<\infty$. A simple interpolating
formula, which can be applied for practical use, will be given in
the next section.

\section{Discussion}
     Let us discuss briefly the higher loop contribution. As was
mentioned in the Sec.\ II, the singularity of $i$-loop analytic
running coupling $^{\text{N}}\alpha^{(i)}_{\text{an}}(q^2)$ at the
point $q=0$ is of the universal type at any loop level. Therefore,
when $q\to0\,$ we have $^{\text{N}}\alpha^{(i)}_{\text{an}}(q^2)\sim
\, ^{\text{N}}\alpha^{\text{(1)}}_{\text{an}}(q^2) \, C^i$, where
$C^i$ are constants. Taking into account that the maximal difference
between $^{\text{N}}\alpha^{(i)}_{\text{an}}(q^2)$ and
$^{\text{N}}\alpha^{\text{(1)}}_{\text{an}}(q^2)$ is in the small
$q^2$ region, we arrive at the following conclusion. The account of
the higher loop corrections leads to changing the slope of the $q\bar
q$ potential $^NV(r)$ when $r\to\infty$. This corresponds to a simple
redefinition of the parameter $\Lambda$ in Eq.~(\ref{NVrAsInf}) at
the higher loop levels.

     As far as the scheme dependence of this approach, we have to
point out the following. It was shown in~\cite{MyDipl} that the
solutions of the analytized RG equation at the higher loop level have
extremely weak scheme dependence. In particular, the solutions of the
RG equation with $\overline{\text{MS}}$ and $\text{MS}$ schemes, are
practically coinciding. Hence, at the higher loop level (there is no
scheme dependence at the one-loop level), the use of different
subtraction schemes leads to the slight variation of the $q\bar q$
potential.

     Thus, neither higher loop corrections, nor scheme dependence,
can affect qualitatively the result obtained in the previous section.

     For the practical use of the new potential it is worth obtaining
a simple explicit expression that approximates it sufficiently well.
For this purpose one can use, for instance, the approximating
function
\begin{eqnarray}
\label{UrDef}
U(r)&=&\frac{8\pi}{3\beta_0}\Lambda\left[
\frac{1}{\ln R}\left(\frac{1}{R}+\frac{R}{2}\right)+
\frac{1}{1-R}\left(\frac{3}{2}+Rf_1(2)\right)\right.\nonumber\\
&&\left.+Rf_1(2)\left(\frac{1}{\ln^2 R}-\frac{1}{(R-1)^2}+
\frac{11}{12}\right)\right],
\end{eqnarray}
which has no any unphysical singularities and possesses the
asymptotics (\ref{NVrAsInf}) and (\ref{NVrAsOrig}). This function is
obtained by smooth sewing the asymptotics
\begin{eqnarray}
\label{NVrAsInfN}
^NV(r)&\simeq&\frac{8\pi}{3\beta_0}\Lambda\cdot\left[
\frac{R}{2\ln(R)}+\frac{Rf_1(2)}{\ln^2(R)}\right],
\quad R \to \infty, \nopagebreak \\
\label{NVrAsOrigN}
^NV(r)&\simeq&\frac{8\pi}{3\beta_0}\Lambda\cdot\frac{1}{R\ln(R)},
\quad R \to 0, \quad R=\Lambda r.
\end{eqnarray}

\begin{figure}
\noindent
\centerline{\epsfig{file=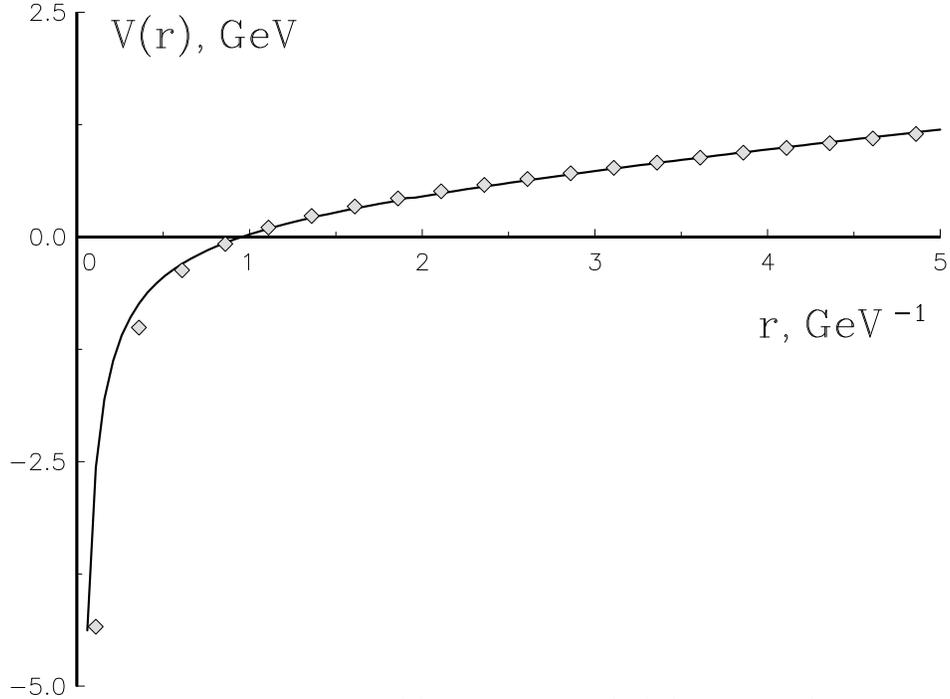, width=125mm}}
\caption{Comparison of the potential $U(r)$ given by
Eq.~(\protect\ref{UrDef}) (solid curve) with the phenomenological
Cornell potential ($\protect\diamond$), Eq.\
(\protect\ref{VrCornell}). The values of the parameters are:
$a=0.39$, $\sigma=0.182$~GeV$^2$ \protect\cite{Brambilla},
$\Lambda=530\,$MeV, $n_f=5$.}
\label{Compare}
\end{figure}

     The formula (\ref{NVrAsInfN}) keeps explicitly the second
leading term of the expansion (\ref{PartInt}), $f_1(2)=0.461$. Some
terms have been introduced into Eq.\ (\ref{UrDef}) only for
eliminating the singularity at the point $R=1$. It should be
mentioned here that the next terms in the expansion (\ref{PartInt})
practically do not affect the shape of $U(r)$. Of course, the
function (\ref{UrDef}) is not the unique interpolating function
between asymptotics (\ref{NVrAsInfN}) and (\ref{NVrAsOrigN}).
Nevertheless, the comparison of $U(r)$ with the phenomenological
potential
\begin{equation}
\label{VrCornell}
^CV(r)=-\frac{4}{3}\frac{a}{r}+\sigma r+\mbox{const}
\end{equation}
(the so-called Cornell potential \cite{Phenom}) shows their almost
complete coincidence (see Fig.\ \ref{Compare}). The fit has been
performed with the use of the least square method in the physical
meaning region $0.1 \leq r \leq 1.0 $~fm \cite{Brambilla}. The varied
parameter in Eq.\ (\ref{UrDef}) is $\Lambda$. The possibility of
shifting the potential $^CV(r)$ in Eq.\ (\ref{VrCornell}) by a
constant was also used. A rough estimation of $\Lambda $ in the
course of this fitting gives $\Lambda \simeq 500$~MeV. This is in
agreement with the values obtained earlier in the framework of the
analytic approach to QCD~\cite{ShSol2}.

\section{Conclusion}
     In the paper the quark-antiquark potential is constructed by
making use of the new analytic running coupling in QCD. This running
coupling arises under analytization of the renormalization group
equation before its solving. The rising behavior of the
quark-antiquark potential at large distances, which provides the
quark confinement, is shown explicitly. The key property of the new
analytic running coupling, leading to the confining potential, is its
infrared singularity at the point $q^2=0$. At small distances, the
standard behavior of the potential, originating in the QCD asymptotic
freedom, is revealed. It is also demonstrated that neither higher
loop corrections, nor scheme dependence, can affect qualitatively the
obtained result. The estimation of the parameter $\Lambda $ in this
approach gives a reasonable value, $\Lambda \simeq 500$~MeV.

     In further studies it would undoubtedly be interesting to
consider in this approach the dependence of the $q\bar q$ potential
on the quark masses.

\acknowledgments
     The author is grateful to Professor D.\ V.\ Shirkov and to Dr.\
I.\ L.\ Solovtsov for valuable discussions and useful comments.
The partial support of RFBR (Grant No.\ 99-01-00091) is appreciated.

\vspace{5mm}

\vspace{5mm}

\end{document}